# Experimental Demonstration of Single Electron Transistors Featuring SiO$_2$ PEALD in Ni-SiO$_2$-Ni Tunnel Junctions

Golnaz Karbasian [a)], Michael S. McConnell, Alexei O. Orlov, Sergei Rouvimov, and Gregory L. Snider

Electrical Engineering Department, University of Notre Dame, IN, USA 46556

[a)] Electronic mail: Golnaz.Karbasian.1@nd.edu

We report the use of plasma-enhanced atomic layer deposition (PEALD) to fabricate single-electron transistors (SETs) featuring ultra-thin (≈1 nm) tunnel-transparent SiO$_2$ in Ni-SiO$_2$-Ni tunnel junctions. We show that as a result of the O$_2$ plasma steps in PEALD of SiO$_2$, the top surface of the underlying Ni electrode is oxidized. Additionally, the bottom surface of the upper Ni layer is also oxidized where it is in contact with the deposited SiO$_2$, most likely as a result of oxygen-containing species on the surface of the SiO$_2$. Due to the presence of these surface parasitic layers of NiO, which exhibit features typical of thermally activated transport, the resistance of Ni-SiO$_2$-Ni tunnel junctions is drastically increased. Moreover, the transport mechanism is changed from quantum tunneling through the dielectric barrier to one consistent with thermally activated resistors in series with tunnel junctions. The reduction of NiO to Ni is therefore required to restore the metal-insulator-metal (MIM) structure of the junctions. Rapid thermal annealing in a forming gas ambient at elevated temperatures is presented as a technique to reduce both parasitic oxide layers. This method is of great interest for devices that rely on MIM tunnel junctions with ultra-thin barriers. Using this technique, we successfully fabricated MIM SETs with minimal trace of parasitic NiO component. We demonstrate that the properties of the tunnel barrier in nanoscale tunnel junctions (with less than $10^{-15}$ $m^2$ in area) can be evaluated by electrical characterization of SETs.

# I. INTRODUCTION

The continuing downscaling of electronic components into the nanoscale calls for reliable and accurate control over fabrication of very thin films down to a few monolayers. The cyclic deposition of atomic layer deposition (ALD), with one monolayer of the desired material deposited in each cycle, makes it a controllable and precise method to form ultrathin tunnel barriers.[1] Plasma enhanced atomic layer deposition (PEALD) has gained significant attention in the past few years since it offers more efficient nucleation, lower required deposition temperature, wider range of available materials for deposition, and better film qualities compared to thermal ALD.[2]

The electronic device of interest is the single electron transistor, a nanoscale electronic device which operates based on the Coulomb blockade phenomenon. An SET consists of a nanoscale island that is separated from source and drain by tunnel transparent, (≈1 nm thick) dielectric barrier. The junction parameters and hence performance of SETs critically depends on the properties and uniformity of ultra-thin tunnel barriers in the nanoscale tunnel junctions.

Here, we report a novel technique to fabricate SETs featuring MIM (Ni-$SiO_2$-Ni) tunnel junctions, where PEALD of $SiO_2$ is used to form an ultra-thin (≈ 1 nm) tunnel-transparent dielectric between two Ni electrodes. However, experiments show that the surface of the bottom Ni electrode is oxidized during the O$_2$ plasma steps in PEALD cycles, forming a layer of NiO that is believed to promote the nucleation of PEALD films on Ni due to the presence of hydroxyl groups on the NiO surface.[3, 4] Additionally, experiments also reveal the formation of NiO at the bottom of the Ni layer evaporated on top of the PEALD SiO$_2$. This phenomenon is attributed to the oxygen containing contaminants on the SiO$_2$ surface[5] that oxidizes a thin layer of the Ni as it is deposited atop the dielectric.

Thin NiO layers on Ni have been reported to exhibit electrical properties consistent with presence of a small (<0.2eV) potential barrier.[6] As a result, the conductance through a few nanometers of NiO is lower than that of Ni by several orders of magnitude at room temperature, and as the device is cooled down, the freeze-out of carriers leads to unmeasurably low conductance. Consequently, the presence of NiO in the tunnel junctions leads to significant performance degradation in the fabricated SETs. We have shown that by annealing in 5% H$_2$-95% Ar ambient at 400°C for 30 minutes following the SiO$_2$ PEALD, the NiO formed on the underlying island can be reduced to Ni. Additionally, the layer of NiO formed on the bottom of the top source/drain leads can be reduced during a second anneal with a similar parameters for 10 minutes. The performance of the fabricated device with the two proposed anneal steps closely matches that of an ideal MIM SET.

Our results show that the study of electron transport in MIM SETs provides a path to characterize the details of formation and modification of the tunnel barrier dielectrics in the junctions.

## II. EXPERIMENT

Since the performance of the SET devices critically depends on the details of electron tunneling in the junctions, trap-free metal-dielectric interface and precise control over the tunnel barrier formation is necessary. To investigate plasma interaction with Ni nanostructures during the barrier formation and to determine the optimal number of PEALD cycles needed to form the dielectric barrier with the desired thickness, two types of test structures were fabricated on a Si wafer covered with 300 nm thermally grown $SiO_2$: (1) a cross-tie device structure, (Fig.1 (a)), forming two Ni-$SiO_2$-Ni tunnel junctions in series, and (2) single layer nanowires covered with PEALD $SiO_2$ (Fig.1 (b)).

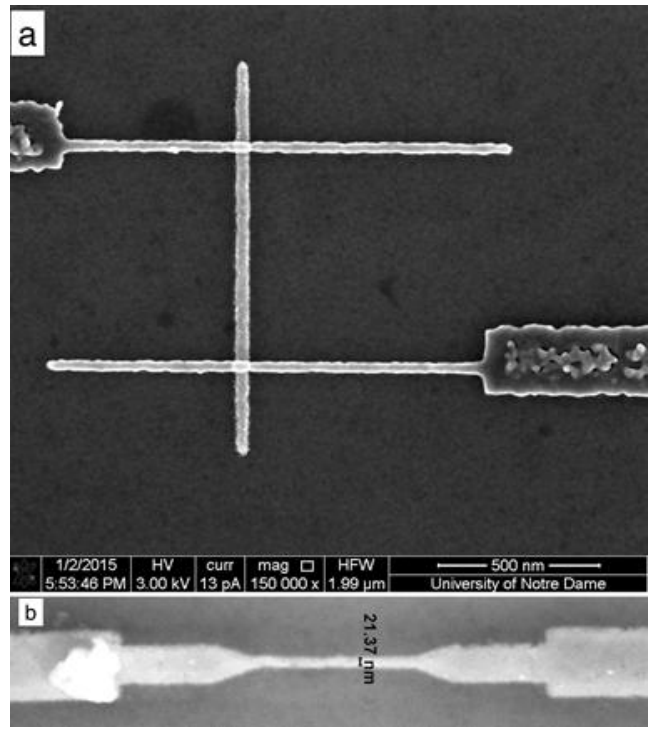

Figure 1 (a) A crosstie test structure composed of two Ni-$SiO_2$-Ni tunnel junctions in series. The differential conductance through the source and drain with respect to the number of $SiO_2$ PEALD cycles in the tunnel junctions was studied.(b) Ni nanowire fabricated on thermal $SiO_2$. The differential conductance of the nanowires was measured prior to and after being covered with PEALD $SiO_2$.

To fabricate the crosstie test structures, a 40 nm wide, 20 nm thick isolated "island" (vertical line in Fig.1 (a)) is first defined using electron beam lithography (EBL) and lift-off. EBL is performed using a 100keV Vistec EBPG 5200 system, while the metal was deposited by e-beam evaporation at base pressure of $<8\times10^{-7}$ Torr at the rate of approximately 0.15 nm/sec. Then, PEALD was performed to form a conformal dielectric film on the island. For $SiO_2$ deposition, with a nominal growth rate of 0.09 nm/cycle, the following PEALD process parameters were used. First, Bis(diethylamino)silane precursor (BDEAS) is introduced in the Oxford FlexAl chamber at 80 mTorr for 125 ms. After the first precursor is purged out of the chamber, 3 seconds of oxygen plasma (pressure of 15 mTorr and power of 300 W) replaces the ligands of the deposited BDEAS monolayer with oxygen. Following the PEALD, 20 nm wide, 30 nm thick source and drain were defined on top of the island by a second EBL, metal deposition, and liftoff.

The 20 nm thick Ni nanowire (Fig. 1(b)), designed to study surface oxidation of Ni during the PEALD process, was defined using EBL and lift off.

Electrical measurements were performed by measuring the differential conductance, $G=dI_{ds}/dV_{ds}$, through both types of structures using standard lock-in techniques at a frequency in the range of 15-80 Hz with AC excitation voltage of 1mV RMS at room temperature and 100 μV at low temperatures. Low temperature measurements are performed using a closed-cycle He refrigerator with minimal temperature of approximately 5K and a closed-cycle $^3$He refrigerator with base temperature of 0.3K; the temperature of the sample is monitored by a thermometer attached to the chip carrier.

For cross-tie devices, reference (Ni-Ni) structures, where the PEALD and therefore the associated oxygen plasma exposure was omitted, were first fabricated and exhibited conductance of G≈0.4 mS between the terminals, which is comparable with the conductance of a nanowire with similar dimensions. However, in devices with only 2 $SiO_2$ PEALD cycles (≈0.2 nm of expected thickness of $SiO_2$ tunnel barrier) the conductance dropped below G≈5 nS. This value is about 4 orders of magnitude smaller than the expected value of G≈ 50 μS, calculated from Simmons approximation for 0.2 nm of uniform $SiO_2$ dielectric barrier in the tunnel junctions.[7] The oxygen plasma step in PEALD processes has been previously reported to oxidize the surface of the underlying substrates.[8-10] Nevertheless, NiO has been shown to be reduced to Ni by annealing the oxidized film in hydrogen at temperatures above 300°C.[11-13] We have found that after annealing at 400°C for 2 min in 5% $H_2$-95% Ar ambient, the fabricated structures with 2 PEALD cycles of $SiO_2$ exhibit a drastic increase in conductance from G ≈5 nS to G> 600 μS.

The lower than expected conductance in the untreated devices with 2 PEALD cycles likely results from NiO layers that are parasitically formed in series with the tunnel barrier during the PEALD process. Note that for the cross-tie structures, the contribution of in series parasitic layer to the total conductance can dominate if its conductance is lower than that of the PEALD barrier: $G_{total}=G_{TJ} \times G_{parasitic}/(G_{TJ}+G_{parasitic}) \approx G_{parasitic}$ if $G_{parasitic} \ll G_{TJ}$.

The anneal in presence of $H_2$ results in reduction of the parasitic NiO layer, and therefore the increase in conductance by 5 orders of magnitude upon annealing the structures in forming gas at 400°C is consistent with hydrogen promoted reduction of the parasitic NiO to Ni. The resulting conductance (G> 600 μS) is much greater than G≈50 μS, expected for a uniform, pinhole-free 0.2 nm of oxide barrier. This is most likely due to non-uniformities in the deposited two monolayers of $SiO_2$, which lead to "pinholes" where the top and bottom Ni electrodes are shorted. Moreover, as discussed below, the increase in Ni grain size caused by the anneal results in reduced electron scattering though the devices,[14] and therefore a higher conductance than that of reference (Ni-Ni) structure is observed.

To estimate the thickness of the NiO layer formed at the Ni-$SiO_2$ interface during the PELAD, we studied the change in the differential conductance along the nanowire (Fig. 1(b)) before and after being covered with 10 cycles of PEALD. After the PEALD process, the Ni nanowires exhibited a decrease in conductance from 1.3 mS to ≈1 mS (the series resistance of the access probes, ≈ 100 Ω, is taken into account in these estimates). The decrease of conductance in nanowires can be interpreted as a decrease in their conducting cross-sectional area due to formation of ≈2 nm NiO on the surface of the wires. Based on our observations, after annealing at 400°C for 30 min in 5% $H_2$-95% Ar ambient, the PEALD $SiO_2$ coated Ni nanowires exhibit an increase in conductance to a level about 30% above that of the as-deposited metal. This result can be interpreted as a cumulative effect of two phenomena: a) reduction of NiO, and b) reduced electron scattering along the wire as a result of the grain size increase caused by the anneal, which results in a higher conductance.[14]

Using the proposed anneal step to reduce the parasitic NiO formed during PEALD on the Ni SET island, we fabricated cross-tie structures in which the island was treated in 5% $H_2$-95% Ar ambient at 400°C for 30 min after being covered with the PEALD tunnel barrier and prior to formation of the source/drain on top of the barrier. Our experiments show more than a 10× increase (from <0.15 nS to 2 nS) in conductance through the structures with 5 cycles of SiO$_2$ barrier. The 2 nS conductance through the device with 5 cycles of SiO$_2$ tunnel barrier and annealed island, however, is still far smaller than that expected based on the Simmons approximation (~1 µS). As a result, we conclude that in addition to the NiO formed on the bottom Ni electrode, there must be another parasitic layer in the tunnel junctions, which results in a lower than expected conductance. To further investigate this assumption, we studied the interface between the top Ni electrode and the PEALD SiO$_2$ using high-resolution transmission electron microscopy (HRTEM).

Figure 2(a) shows an HRTEM image of 40 cycles of PEALD SiO$_2$ between two Ni films. A relatively thick SiO$_2$ is deposited to improve the identification of NiO at the top interface. Figure 2(b) shows the elemental composition and concentration in 0.1 nm spaced data points along the red arrow in Fig. 2(a) extracted from energy dispersive X-ray spectroscopy (EDX). The points 2-6 denote the area near the top Ni-SiO$_2$ interface where the presence of NiO can be inferred from the pronounced oxygen content in the vicinity of the top Ni electrode that is more than twice the Si concentration, for a stoichiometric SiO$_2$. It should be mentioned that the maximum 20% concentration of Ni is due to the dominating presence of copper (≈80%) resulting from the copper grid that is supporting the sample in the TEM system.

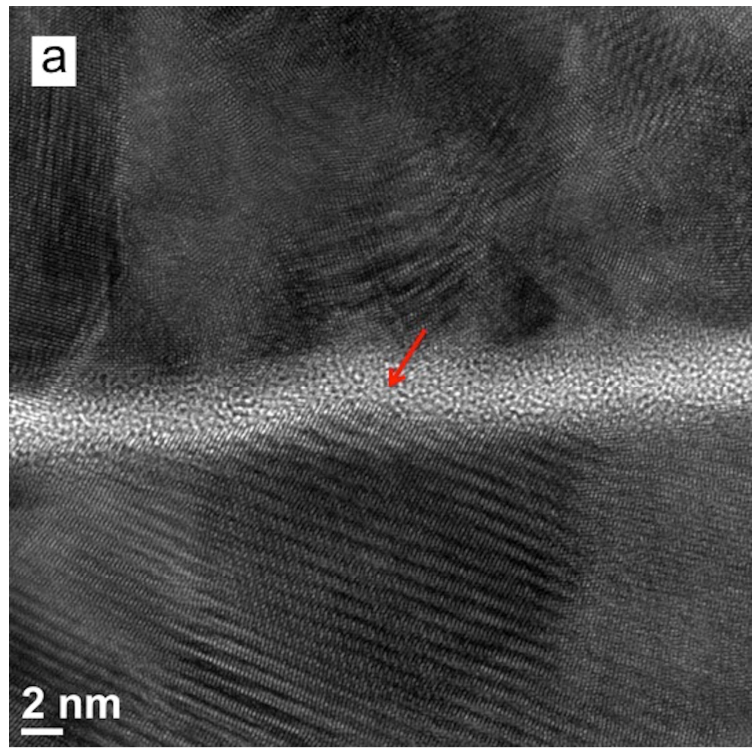

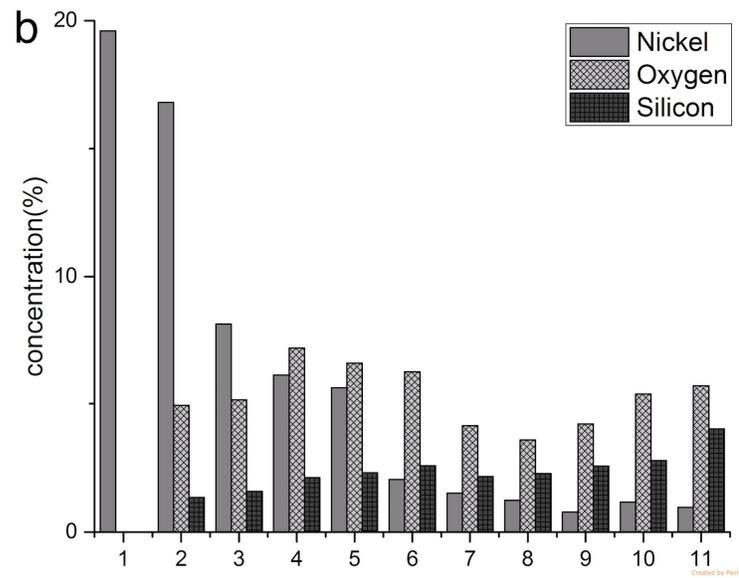

Figure 2 (Color online) (a) The TEM image of a Ni-SiO$_2$-Ni structure fabricated with 40 cycles of PEALD SiO$_2$. (b) Elemental composition along the red arrow in (a): at the top Ni-SiO$_2$ interface, oxygen concentration more than twice the Si concentration suggests the presence of NiO.

The main cause of oxidation in the top Ni remains to be investigated, but the oxidation of metals when deposited on $SiO_2$ has been reported for Cu, Mo, and W and has been attributed to the oxygen containing contaminants, such as water, on the oxide surface.[15, 16]

Based on our observations, all of the crosstie devices with 2, 5, 8, and 10 cycles of PEALD $SiO_2$ as the barrier exhibit a conductance greater than 400μS when prior to formation of the source/drain on the PEALD $SiO_2$ covered island, the NiO layer formed on the island is reduced for 30 min at 400°C in 5% $H_2$-95% Ar and the top NiO, formed on the bottom of the source/drain, is reduced in the same conditions for 10 min after the source/drain are defined. In contrast, after the two reducing treatments, the average conductance for devices with 12 or 13 cycles of $SiO_2$ as the tunnel barrier is 0.1 μS, which is consistent within an order of magnitude of the expected tunneling conductance through a ≈1.1 nm of $SiO_2$ barrier.[7] We attribute these results to non-uniformities in the PEALD films that are more detrimental for thinner dielectrics, and lead to a high conductance after the parasitic NiO layers in the junctions are reduced.

To verify the validity of oxidation/reduction of NiO layers at Ni/$SiO_2$ interfaces, annealing of the fabricated cross-tie samples in Ar plasma at 300°C has also been performed and no significant increase in conductance was observed through the structures. This confirms the dominant effect of $H_2$ in reducing the NiO and increasing the conductance in these devices. Figure 3 shows the schematic of dielectric barrier formation in Ni-$SiO_2$-Ni tunnel junctions, illustrating where the parasitic NiO layers are formed and then reduced during the tunnel junction fabrication.

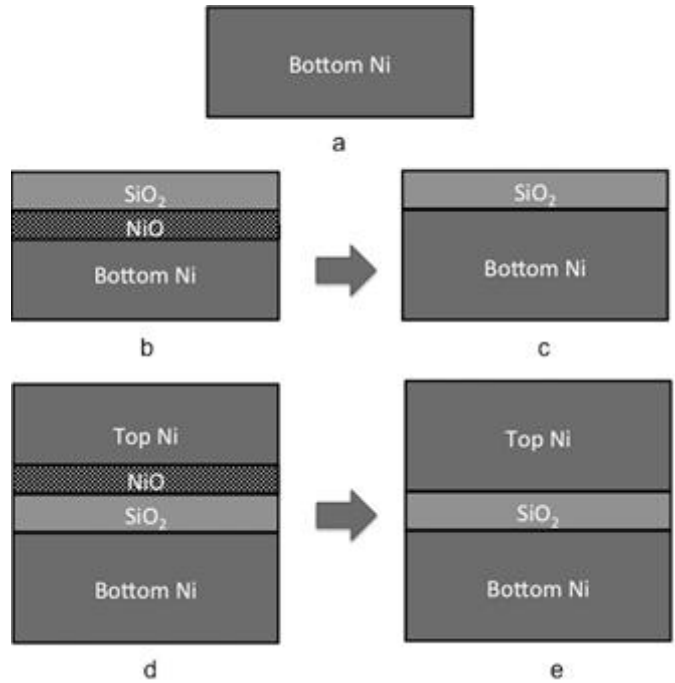

Figure 3 Schematic of parasitic NiO formation during fabrication of Ni-SiO$_2$-Ni tunnel junctions. (a) The Ni bottom electrode is evaporated on the substrate. (b) The PEALD of SiO$_2$ forms the tunnel barrier on the bottom Ni electrode. Additionally, NiO is formed on the surface of the bottom Ni due to the oxidizing effect of the O$_2$ plasma steps in PEALD of SiO$_2$. (c) The underlying NiO is reduced when the SiO$_2$ covered Ni is annealed in H$_2$ containing environment at 400°C for 30 minutes, while the SiO$_2$ layer remains intact. (d) Ni top electrode is evaporated atop the SiO$_2$. The bottom of the top Ni electrode is oxidized due to residual oxygen containing contaminants on the SiO$_2$. (e) The parasitic NiO on the top layer is reduced when the fabricated structure is annealed in H$_2$ containing environment at 400°C for 10 minutes.

## III. Ni-SiO$_2$-Ni TUNNEL JUNCTIONS FOR METALLIC SETS

From the fabrication of cross-tie devices it was concluded that 12-13 PEALD cycles of SiO$_2$ is suitable for Ni-SiO$_2$-Ni tunnel junctions in SET devices since the resulting conductance is sufficiently below the quantum conductance, $G_q \approx 40$ μS, required for suppression of quantum fluctuations in single electron transistors,[17] yet high enough (>10 nS) to avoid signal-to noise ratio problems due to excess Johnson noise.

In order to further decrease the junction capacitance between the island and the source and drain leads, $C_d$ and $C_s$, and thus increase the operating temperature of the fabricated devices, we decreased the junction area by fabricating the island inlaid in the oxide substrate. This reduces the island area accessible to the source and drain from three sides of the island to just the top of the island. To fabricate the inlaid island devices, the following steps were performed: first, the pattern of the island is formed in polymethylglutarimide (PMGI) spun on the thermal $SiO_2$ substrate. PMGI can be used as a mask for $SiO_2$ etch due to its higher etch resistance than polymethylmethacrylate (PMMA)[18]. The pattern of the island is then etched into the oxide using Ar, $C_4F_8$, $CHF_3$, and $CF_4$ chemistry in an inductively coupled plasma (ICP) etcher. Next, Ni is deposited by e-beam evaporation to fill the trench in $SiO_2$, while the field is still covered by the PMGI mask. The evaporated metal on the field along with the underlying PMGI is then lifted off the sample by MR-Rem 400 from Micro Resist Technology, heated to 70°C. Chemical mechanical polishing (CMP) is then used to remove any residual Ni on the oxide field, while leaving the island trench filled with Ni. Next, a $SiO_2$ PEALD process is performed to form a dielectric barrier covering the island, followed by annealing in a 95% Ar-5% $H_2$ ambient at 400°C for 30 min to reduce the NiO on the bottom electrode. Finally, Ni source and drain are defined by a second EBL and liftoff, and the whole structure is annealed for 10 minutes in a 95% Ar-5% $H_2$ ambient at 400°C so the NiO formed on the bottom of the source and drain electrodes is reduced.

It is worth mentioning that since the size of the island depends on the profile of the trench etched in the oxide, one can design the plasma etch process to achieve a trench in the substrate in which the bottom of the trench is narrower than its opening. The CMP step can then be tuned to thin down the island and leave a wire that is narrower than the opening in the PMGI mask. The described technique to form the cross-tie structures results in at least 2× reduction of the overlap area for the two crossing wires that form the tunnel junctions. The use of CMP can also enable a full-damascene metallic SET.[19]

Figure 4 shows the conductance $G_{ds}$ at 5.4 K plotted as a function of the applied bias $V_{ds}$, for two devices: "treated" sample, (solid line) with 13 $SiO_2$ PEALD cycles forming the tunnel barrier in which the bottom NiO layer is reduced for 30 min at 400°C in 5% $H_2$ -95% Ar and the top NiO is reduced at the same conditions for 10 min (solid black curve), and "untreated" sample with 12 $SiO_2$ PEALD cycles (dashed line) for which source/drain leads were not treated with $H_2$. To get a comparable scale, the conductance for each device is normalized by its high temperature, $T \gg E_C/k_B$, value ($G_{high-T}$), where $E_C$ is charging energy and $k_B$ is the Boltzmann constant.

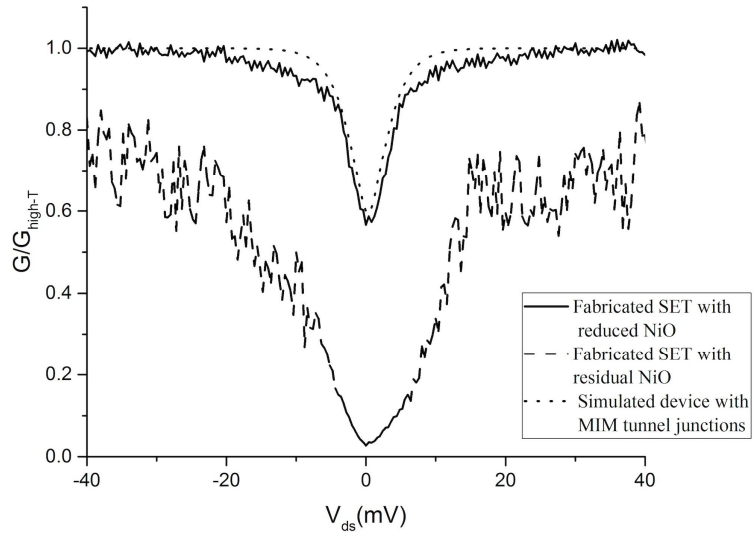

Figure 4. $G_{ds}(V_{ds})$ at 5.4 K for a "treated" sample (solid line) in which the island is annealed in 5%$H_2$-95%Ar at 400°C for 30 minutes, and the finished device is annealed for 10 min in similar conditions after source and drain are formed on the PEALD $SiO_2$-covered island. The dashed line represents an "untreated" sample with untreated source/drain leads. The dotted line denotes an SET simulated based on orthodox Coulomb blockade theory for a metal dot. The parameters of the junctions are extracted from the charging diagram of the device at 0.4 K (Fig 5 (b)).

According to orthodox model of Coulomb blockade[17], when an MIM SET is cooled to T≈ $E_C/k_B$, the conductance at $V_{ds}$=0 decreases towards $G_{high-T}/2$ due to a reduction in the island occupation probability imposed by Coulomb blockade[20]. For $|V_{ds}|>$ $2E_C/e$, however, the Coulomb blockade is suppressed and SETs exhibit an almost constant differential conductance that approaches the high temperature value $G_{ds}/G_{high-T}$=1. The treated sample in Fig. 4, with two reducing anneal steps, clearly exhibits this type of $G_{ds}(V_{ds})$ dependence. In contrast, in the untreated sample (the dashed line in Fig. 4), the conductance is strongly suppressed at $V_{ds}$=0, and gradually increases with applied bias. This behavior is consistent with presence of parasitic NiO layer in series with tunnel junctions (i.e. [Ni-NiO]-$SiO_2$-[NiO-Ni] instead of Ni-$SiO_2$-Ni), as it modifies the overall barrier for electron transport[6]. This experiment confirms that the parasitic NiO is responsible for the activated conductance in the device structure, and shows that the proposed double-anneal procedure in 5%$H_2$-95% Ar ambient at 400°C is effective in reducing the NiO parasitic layers that are formed at the interfaces of the tunnel junctions.

Figure 5(a) shows the SEM micrograph of the treated sample and Fig. 5 (b) shows the charge stability diagram, also referred to as Coulomb diamonds, for the same device obtained at T=0.4K. The simulations of the charging diagram, Fig 5(c), based on orthodox Coulomb blockade theory for a metal dot,[21] with junction parameters of $C_s$=40 aF, $G_s$= 0.25 µS, $C_d$=68.4 aF, $G_d$= 0.15 µS, and $C_g$=1.3 aF best matches the experiments, where the values of the junction capacitances are extracted from the slopes of the experimentally measured diamonds, $\gamma^+=C_g/C_g+C_s$, $\gamma^-= -C_g/C_d$, and the island-gate capacitance, $C_g$, can be calculated from the period of the diamonds ($\Delta V_g$): $C_g=e/\Delta V_g$=1.3 aF.

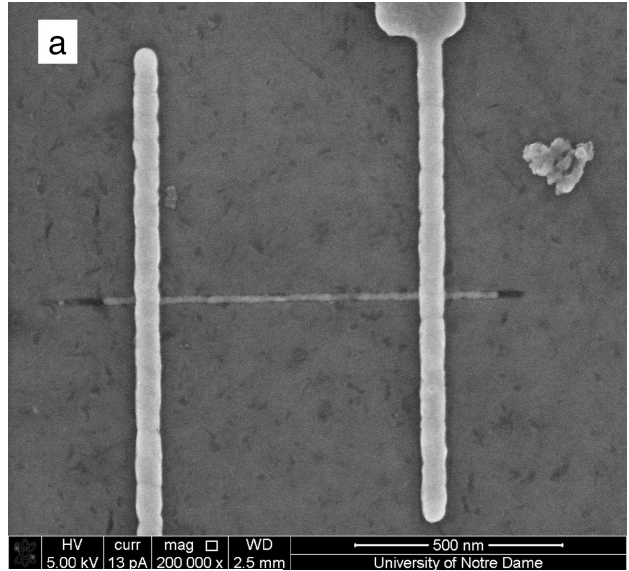
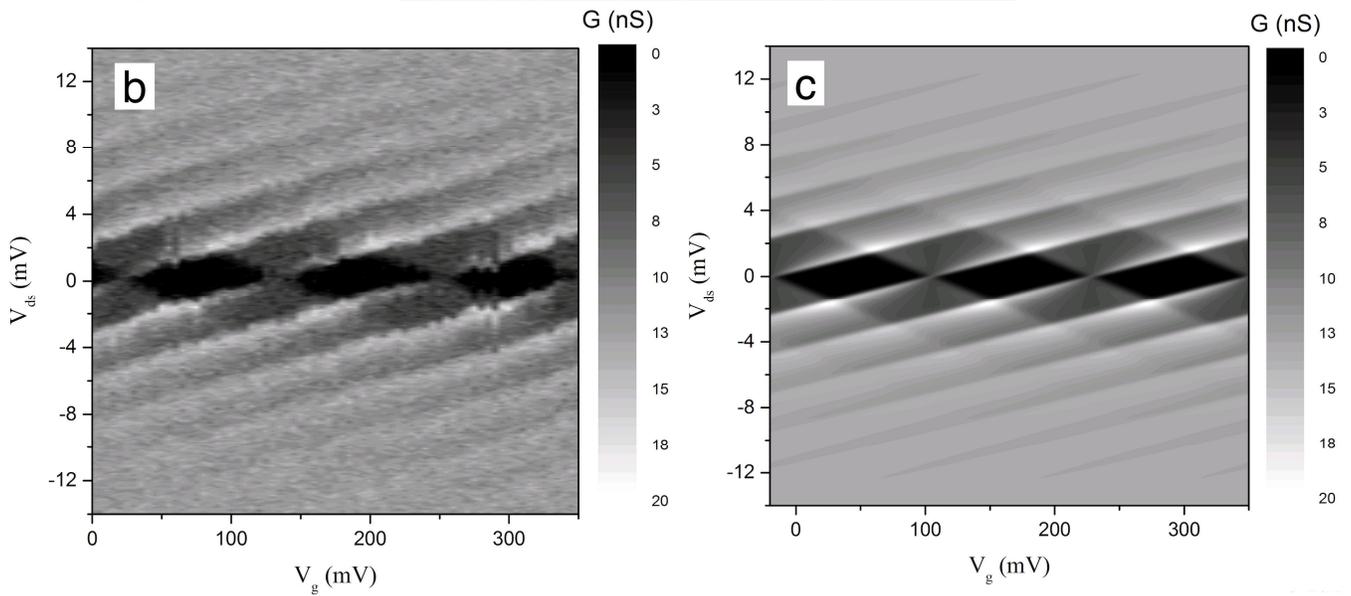

Figure 5 (a) SEM image of a treated single electron transistor fabricated with Ni-SiO$_2$-Ni tunnel junctions. (b) Charging diagram (Coulomb diamonds) of the fabricated device at 0.4 K. (c) A simulated charging diagram of a metallic SET with $C_s$=40aF, $G_s$= 0.25 μS, $C_d$=68.4 aF, $G_d$= 0.15 μS, and $C_g$=1.3 aF at T=0.4K.

The dotted line in Fig. 4 represents the calculated $G_{ds}(V_{ds})$ at 5.4 K using the extracted parameters; a good correlation between the theory and experiment is observed. This confirms the formation of well-defined MIM tunnel junctions in the fabricated device and proves the effect of the anneal steps in elimination the NiO parasitic layers by means of reduction.

Based on the dimensions of the imaged SET in Fig. 5(a), the expected junction capacitance values are estimated using parallel-plate capacitor approximation: $C_d = C_s = \varepsilon_o \varepsilon A/d \approx 20$ aF where $\varepsilon_o$ and $\varepsilon$ are the vacuum permittivity and the dielectric constant of $SiO_2$, respectively, while A is the junction area, $46 \times 13$ nm$^2$, and d=1.17 nm is the nominal thickness of 13 cycles of $SiO_2$ tunnel barrier. The observed discrepancy (2 to 3.5 times larger values of the experimentally observed capacitances) as well as dissimilarities in capacitance values for the junctions can be attributed to thinner than expected dielectric layer, perhaps due to a delay in growth during the first few PEALD cycles that results in an average growth rate slower than the nominal 0.09 nm/cycle, non-uniformities of $SiO_2$ formation, or intermixing of $SiO_2$ and Ni at the interfaces. Further study regarding this discrepancy is in progress.

## IV. CONCLUSIONS

We present a technique for fabrication of metal-insulator-metal single-electron transistors featuring an ultra-thin (≈1 nm) $SiO_2$ tunnel barrier between two Ni electrodes using plasma enhanced atomic layer deposition. We demonstrate that the $O_2$ plasma in the PELAD process is responsible for formation of thin layer of NiO at the Ni-$SiO_2$ interface. In addition to the oxidized upper surface on the bottom Ni electrode, parasitic

NiO is formed at the bottom of the upper Ni electrode, most likely due to the oxidizing effect of the oxygen containing contaminants such as water adsorbed on the $SiO_2$ layer when this layer is exposed to the ambient. The presence of these two parasitic layers in the tunnel junctions (i.e. [Ni-NiO]-$SiO_2$-[NiO-Ni] instead of Ni-$SiO_2$-Ni) significantly increases the resistance through the junctions and changes the conduction mechanism from direct tunneling to a combined tunneling-activated carrier transport which at low temperature leads to freeze out of conductance through the device. We proposed and implemented a technique to reduce both parasitic NiO layers by using two separate anneal steps in 5% $H_2$-95% Ar ambient at 400°C. Using the proposed process, we fabricated single electron transistors with Ni-$SiO_2$-Ni tunnel junctions that show the electrical characteristics expected for devices with MIM junctions. A good correlation between the orthodox model and experiment is observed, which shows that the proposed reducing step in effective in reducing the NiO parasitic layers that are formed in the tunnel junctions.

## ACKNOWLEDGMENTS

This work was supported by National Science Foundation grants DMR-1207394 and CHE-1124762. The authors are grateful to Dr. Alexander Mukasyan for multiple useful discussions.